\documentclass[a4paper]{jpconf}
\usepackage{amssymb}
\usepackage{amsthm}
\usepackage{latexsym}
\usepackage{graphicx,epsfig}
\def\beq{\begin{equation}}
\def\eeq{\end{equation}}
\def\br{\begin{eqnarray}}
\def\er{\end{eqnarray}}
\def\bfg{\begin{figure}}
\def\efg{\end{figure}}
\def\bit{\begin{itemize}}
\def\eit{\end{itemize}}

\def\pa{\partial}
\def\l{\left}
\def\r{\right}

\def\nn{\nonumber}

\def\a  {\alpha}
\def\b  {\beta}
\def\c  {\gamma}

\def\d  {\delta}

\def\L  {\Lambda}

\def\n  {\nu}
\def\o  {\omega}
\def\O  {\Omega}
\def\p  {\pi}

\def\t  {\tau}
\def\vp{\varphi}
\def\la {\label}
\def\pa {\partial}
\def\ba{\begin{eqnarray}}
\def\ea{\end{eqnarray}}
\def\f {\frac}

\def\bi {\begin{itemize}}
\def\ei {\end{itemize}}

\def\ub {\underbrace}
\def\le {\left}
\def\ri {\right}
\def\be {\begin{equation}}
\def\ee {\end{equation}}
\def\la {\label}

\def\benu{\begin{enumerate}}
\def\eenu{\end{enumerate}}
\def\nn{\nonumber} 
\def\pa{{\partial}}
\def\l{\left}
\def\r{\right}

\newcommand{\SBH}{S_{_{\rm BH}}}
\newcommand{\SEN}{S_{_{\rm ent}}}

\usepackage{color}

\def\blue{\textcolor{blue}}
\def\black{\textcolor{black}}

\begin{document}

\title{Power-law corrections to black-hole entropy via entanglement}
\author{Saurya Das${}^{\dagger}$\footnote{Email: saurya.das@uleth.ca}, 
S. Shankaranarayanan${}^{\ddagger,\star}$\footnote{Email: Shanki.Subramaniam@port.ac.uk}
Sourav Sur${}^{\dag}$\footnote{Email: sourav.sur@uleth.ca}}
\address{${}^{\dagger}$Dept. of Physics, University of Lethbridge,\\
4401 University Drive, Lethbridge, Alberta, Canada T1K 3M4 \\
${}^{\ddagger}$Max-Planck-Institut f\"ur Gravitationphysik,
Am M\"uhlenberg 1, D-14476 Potsdam, Germany \\
${}^{\star}$Institute of Cosmology and Gravitation, 
U. of Portsmouth, Portsmouth PO1 2EG, UK}

%\maketitle

\begin{abstract}
We consider the entanglement between quantum field degrees of freedom
inside and outside the horizon as a plausible source of black-hole
entropy. We examine possible deviations of black hole entropy from
area proportionality. We show that while the area law holds when the
field is in its ground state, a correction term proportional to a
fractional power of area results when the field is in a superposition
of ground and excited states. We compare our results
with the other approaches in the literature.
\end{abstract}

\section{Introduction}

It is now well-known that the black-hole (commonly referred to as
Bekenstein-Hawking) entropy is finite and is given by the relation 
\cite{Bekenstein:1973ur}
\beq
\SBH = \frac{k_B}{4} \frac{A_{_H}}{\ell_P^2}  \qquad {\rm where} \qquad
\ell_P^2 \equiv \frac{\hbar G}{c^3} \, , 
\eeq
$A_{_H}$ is the area of the black-hole horizon. It is infinite, if
both the matter and gravity are purely classical! In other words, the
finiteness of the black-hole entropy (and the validity of second-law of
thermodynamics) requires that the matter and(or) gravity to be
quantized.

Although the above relation has been derived by various approaches 
\cite{stringsetc,Bombelli:1986rw,Srednicki:1993im,Das:2005ah}, we do
not yet understand:
Why $\SBH$ is not extensive?  
What is the microscopic origin of black-hole entropy? 
Are there corrections to the Bekenstein-Hawking entropy?
and  How generic are these corrections? 

The purpose of this paper is an attempt to understand the generic
corrections to the Bekenstein-Hawking entropy by assuming entanglement
as a source of black-hole entropy. We show that while the area law
holds when the field is in its ground state, a correction term
proportional to a fractional power of area results when the field is
in a superposition of ground and excited states.

The paper is organized as follows: In the next section, we briefly
discuss various properties of entanglement entropy and provide
motivation for the relevance of the entanglement entropy to $\SBH$. In
Sec. (\ref{sec:3}), we discuss the setup and the key steps in
evaluating the entanglement entropy of scalar field propagating in
flat space-time. In Sec. (\ref{sec:4}), we obtain the entanglement
entropy for a superposed state and show explicitly that the power-law
corrections become important for small horizon limit. We conclude with
a summary in Sec. (\ref{sec:conc}). In appendix, we show explicitly
that at a fixed Lemaitre time, the Hamiltonian of a scalar field in a
black hole background reduces to that in flat space-time.

%%%%%%%%%%%%%%%% NEW SECTION
\section{Entanglement entropy}
\label{sec:2}

Let us consider a bipartite quantum system ${\mathcal I} \equiv 
\{1 2\}$ and assume that ${\mathcal I}$ 
is in a pure state $|\Psi \rangle$. The entanglement entropy
($\SEN$) is defined as
\beq
\SEN(\Psi) \equiv S(\rho_1) = S(\rho_2)
\label{eq:2}
\eeq
where 
\beq
S(\rho_1) = - Tr[\rho_1 \ln(\rho_1)] = - \sum_{n} p_n \ln p_n 
\label{eq:VNent}
\eeq
is the Von Neumann entropy, $p_n$ are the Eigen-values of $\rho_1$ 
obtained by solving the integral equation
\beq
\int_{-\infty}^{\infty} dx' \rho_1(x,x') f_{n}(x') = p_n f_{n}(x)
\label{eq:integraleq}
\eeq
and 
\beq
\rho_1 = Tr_{2}(|\Phi \rangle \langle \Phi|)
\eeq
is the reduced density matrix obtained by tracing over 2. 
Note that the entanglement entropy (\ref{eq:2}) is independent of the 
which degrees of freedom are traced over.

In order to set the ideas, we consider a simple example --- 
2-coupled harmonic oscillator. The Hamiltonian of a coupled 
Harmonic oscillator is given by 
\beq
H = \f{1}{2} \l[ p_1^2 + p_2^2 + k_0 \l( x_1^2 
+ x_2^2 \r) + k_1 \l( x_1 - x_2\r)^2 \r] 
\eeq
where $(x_i, p_i)$ are coordinate and momenta of the oscillator,
respectively, and $k_1$ is the interaction term which is assumed to be
positive. The quantization of the above Hamiltonian is straight
forward in the normal coordinates:
\beq
x_\pm = \f{x_1\pm x_2}{\sqrt{2}} \, . 
\eeq
The ground state wave function is given by
\beq
\psi_0 \le( x_1, x_2\ri) 
= \frac{\l( \o_+\o_- \r)^{1/4}}{\p^{1/2}}
\exp\l[- \le(\o_+ x_+^2 + \o_- x_-^2\r)/2\r] \nn 
\eeq
where 
\beq
\o_+ = \sqrt{k_0};~~ \o_- = \sqrt{k_0+2k_1};~~ R^2 =
\frac{\omega_+}{\o_-} < 1 \, .
\eeq
\begin{figure}[!hbt]
\begin{center}
\includegraphics[width=9.5cm,height=6.75cm]{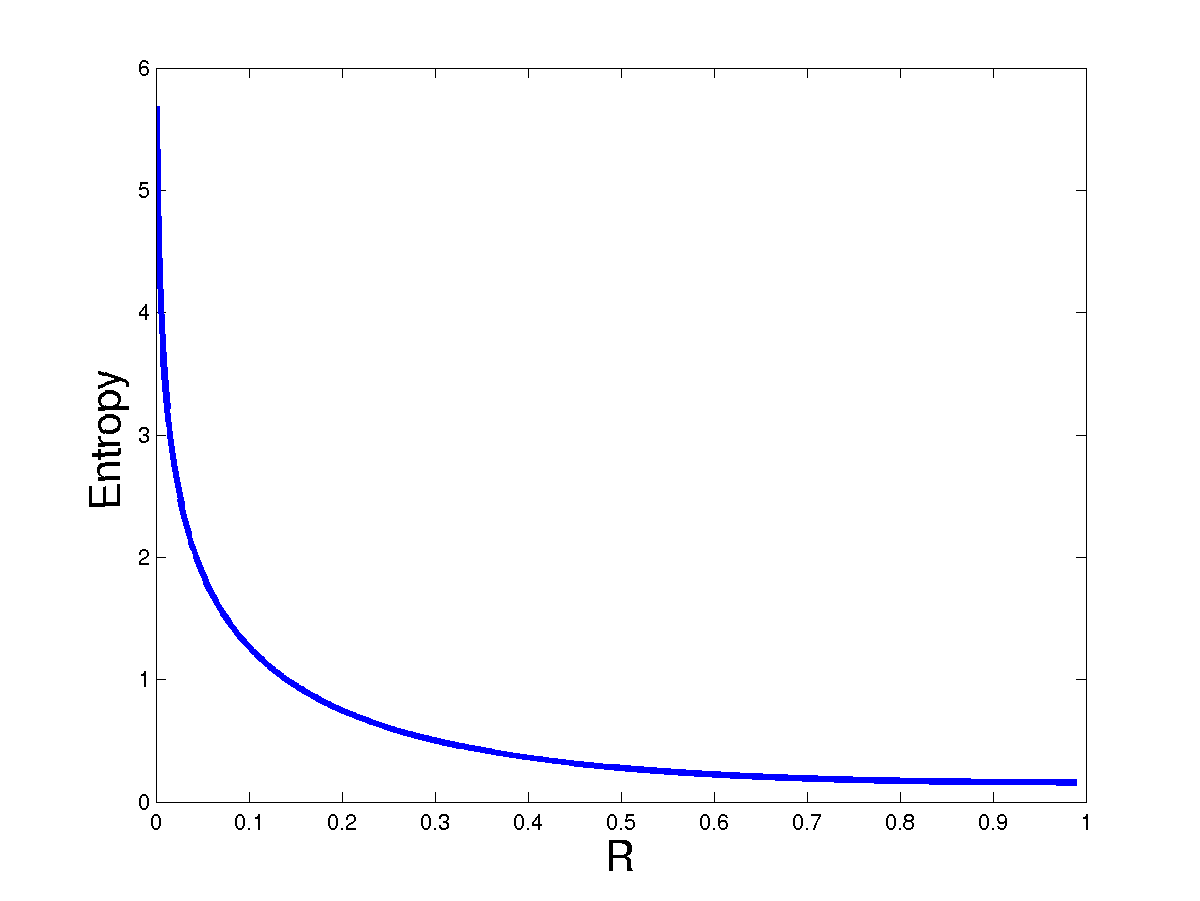} 
\caption{Plot of entanglement entropy of a coupled harmonic oscillator 
and $R$.}
\end{center}
\label{fig:1}
\end{figure}
Let us consider the situation in which we choose not to have any
information about the oscillator $1$. Mathematically, this corresponds
to tracing over the oscillator 1, i. e.,
\br
\rho_{2} \l( x_2, x_2'\r) 
&=&  \int_{-\infty}^\infty dx_1 \psi_0 \l( x_1, x_2\r)
\psi_0^\star \l( x_1, x_2'\r)  \nn \\
%%%
& = & \sqrt{\f{\c-\b}{\p}}
\exp\l[ -\frac{\c}{2}\l( x_2^2+x_2'^2 \r) +\b x_2 x_2'  \r] 
\label{eq:HO-denMat}
\er
where 
\beq
\b=\f{\o_-(1-R^2)^2}{4(1+R^2)}; \quad
\gamma = \o_{-} \f{1+6R^2+R^4}{4(1+R^2)}; \quad 
\xi = \l(\frac{1 - R}{1 + R}\r)^2 < 1 \, .
\eeq
Substituting (\ref{eq:HO-denMat}) in the integral equation 
(\ref{eq:integraleq}), the eigenfunctions and eigenvalues are 
given by
\beq
f_n(x) = H_n\le(\sqrt{\a} x \ri) \exp\le( - \frac{\a x^2}{2}\ri) 
\qquad  p_n = \le(1 -\xi\ri) \xi^n~ \, .
\eeq
Substituting the eigenvalues in (\ref{eq:VNent}), the entanglement 
entropy is given by 
\br
\SEN(\xi) = - \ln(1 - \xi) -  \frac{\xi}{1- \xi} \ln(\xi) \, .
\er

In Fig. (\ref{fig:1}), we have plotted $\SEN$ of the coupled
oscillator in terms of $R$. It is interesting to note that 
the entanglement entropy vanishes when the oscillators are 
uncoupled ($R = 1$) while they diverge in the strongly 
coupled limit ($R = 0$).

But what has this digression to do with black-holes and what is the
relevance of $\SEN$ to black-hole entropy? This can be
understood from the fact that both entropies are associated with the
existence of the horizons. Consider a scalar field on a background of
a collapsing star. At early times, there is no horizon, and both the
entropies are zero. However, once the horizon forms, $\SBH$ is
non-zero, and if the scalar field degrees of freedom inside the
horizon are traced over, $\SEN$ obtained from the reduced density
matrix is non-zero as well.

In the next section, we give the procedure for evaluating the 
entanglement entropy of quantum scalar field propagating in 
flat space-time. In the Appendix, we have shown explicitly that at a
fixed Lemaitre time, the Hamiltonian of a scalar field in 
Schwarzschild space-time reduces to that in flat space-time. 
All the results we derive in case of flat space-time can then be 
extended to the black-hole space-time at a fixed Lemaitre time.

%%%%%%%%%%%%%%%% NEW SECTION
\section{Entanglement entropy for scalar fields -- Setup}
\label{sec:3}

\begin{figure}[!htb]
\begin{center}
\includegraphics[width=7.5cm,height=5.7cm]{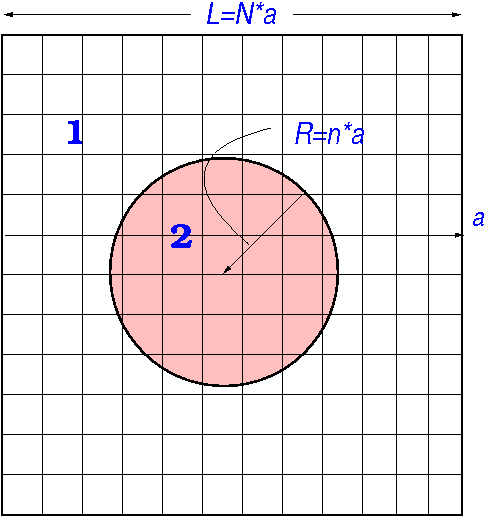}
\end{center}
\caption{Sketch of the discretization of the scalar field propagating 
in the flat space-time. The shaded region is one we trace over.}
\label{fig:2} 
\end{figure}

The Hamiltonian of a free massless scalar field propagating in the
Minkowski space-time is given by
\beq
H = \f{1}{2} \int d^3x \le[ \pi^2(\vec r) + |\nabla \varphi(\vec
  r)|^2\ri]  \, .
\label{ham2}
\eeq
For simplicity, let us assume that the scalar field in confined in a
spherical box $\{1 2\}$. The cartoon version of the setup is provided
in Fig (\ref{fig:2}).
Partial-wave decomposition of the scalar field and its canonical 
conjugate
\br
\varphi_{lm} (r)  = r \int d\O~Z_{lm} (\theta,\phi)
\varphi (\vec r); ~~ 
\p_{lm}(r) = r \int d\O~Z_{lm} (\theta,\phi)  \p (\vec r)
\er
leads to the following reduced Hamiltonian:
\beq
H = \sum_{lm} H_{lm} = \sum_{lm}
\f{1}{2} \int_{0}^{\infty} \!\!\!\! dr 
\le[ \pi_{lm}^2(r) + r^2  
\l[\frac{\pa}{\pa r}\l(\frac{\varphi_{lm}}{r}\r)\r]^2 
+ \frac{l (l + 1)}{r^2}\varphi_{lm}\ri] \, ,
\label{eq:red-Ham}
\eeq
where $Z_{lm}$ are the real Spherical harmonics. (For details,
see Refs. \cite{Das:2005ah,Das:2006vx,Das:2007sj}.)

The computation of the entanglement entropy involves three steps:
(i) Discretize the Hamiltonian, 
(ii) Choose a quantum state, 
and (iii) Trace over region 2 (or 1) to obtain the density
matrix.  

Note that, even if we obtain closed-form expression of the density
matrix, it is not possible to analytically evaluate the entanglement
entropy. Hence, we need to resort to numerical methods. In the rest of
the section, we discuss these steps in detail.

\subsection{Discretized Hamiltonian}

Discretizing the Hamiltonian (\ref{eq:red-Ham}) in a spherical lattice
as described in Fig. (\ref{fig:2}), we get
\br
H_{lm}&=& \!\! 
\f{1}{2a} \sum_{j=1}^N \le[ \p_{lm,j}^2 + \le(j+\f{1}{2}\ri)^2
\le( \f{\varphi_{lm,j}}{j} - \f{\varphi_{lm,j+1}}{j+1} \ri)^2 \r. 
+ \l. \f{l(l+1)}{j^2}~\varphi_{lm,j}^2 \ri] \nn  \\
&=& \f{1}{2a} \sum_{j=1}^N \p_{j}^2 
+ \f{1}{2a} \sum_{i,j=1}^N \vp_i K_{ij} \vp_j \, ,
\label{eq:Ncoupl-HO}
\er
where $a$ is the lattice spacing, $L$ ($\gg a$) is the length of the
box, $N$ is the number of lattice points and the interaction 
matrix $K_{ij}$ is given by 

\begin{minipage}{8.5cm}
\br
\!\!\!\!\!\!\!\!
K_{ij} &=& \frac{1}{i^2} 
\le[ 
\le(i+\f{1}{2}\ri)^2 + \le(i - \f{1}{2} \ri)^2
\ri] \d_{i,j(i\neq 1,N)} \nn\\
%%%%
\!\!\!\!\!\!\!\!
& & 
{ \frac{1}{i^2} \le[l(l+1) + 
\f{9}{4}~\d_{i1} \d_{j1} 
+ \le[ N - \f{1}{2}\ri]^2 \d_{iN} \d_{jN} \r]
} \nn \\ 
%%%%
\!\!\!\!\!\!\!\!
&&  
\ub{\blue{-\le[\f{(j+\f{1}{2} )^2}{j(j+1)} 
\ri] \delta_{i,j+1}
- \le[
\f{(i+\f{1}{2} )^2}{i(i+1)} 
\ri] \delta_{i,j-1}} }_{\black{\mbox{nearest neighbor interaction}}}  \nn
\er    
\end{minipage}
%%%%
\begin{minipage}{4.4cm}
\br
{K}  = \le( \begin{array}{lllllll} 
{\times} & \blue{\times} & {}& {}& {} & {} & {} \\
\blue{\times} & {\times} & \blue{\times} & {}& {} & {} & {} \\
{}  & \blue{\times} & {\times} & \blue{\times} & {} & {} & {} \\
{} & {}  & \blue{\times} & {\times} & \blue{\times} & {} & {} \\
{} & {} & {}  & \blue{\times} & {\times} & \blue{\times} & {}  \\
{} & {} & {} & {}  & \blue{\times} & {\times} & \blue{\times}  \\
{} & {} & {} & {} & {}  & \blue{\times} & {\times}  
%{} & {} & {} & {} & {} & {}  & {\times}   
\nn \\[8pt]
\end{array} \ri) 
\er
\end{minipage}
\vspace*{6pt}

Note that the Hamiltonian (\ref{eq:Ncoupl-HO}) is a 
$N$-coupled Harmonic oscillators which satisfy the following 
commutation relations:
\beq
[\vp_{lm,j}, \p_{l'm',j'}] = i \delta_{l l'} \delta{m m'} 
\delta{j j'}
\eeq

\subsection{Choice of quantum state}

The general Eigen state of the $N$-coupled harmonic oscillators is
\beq
\Psi (x_1,\dots,x_n,x_{n+1},\dots,x_N) =
\le[\f{|\O|}{\p^N}\ri]^{1/4} 
~ \exp\le[ -\f{x^T\cdot \O \cdot x}{2} \ri] 
\times
\prod_{i=1}^{N} 
\frac{1}{\sqrt{2^{\nu_i} \nu_i}}~
H_{\n_i} \le( {K_D^{\f{1}{4}}}_i~\bar{x}_i\ri)
\nn 
\eeq
where $K_D$ is the diagonal matrix of $K$ obtained by the unitary
transformation $K = U^{T} K_D U$, $U$ is a unitary matrix and $\Omega
= U^{T} K_D^{1/2} U$. 

It is {\it not possible} to obtain a closed form expression for the
density matrix for such a general state. To make the calculations 
tractable, we make two choices for the $N$-particle state:
\begin{enumerate}
\item {\it Vacuum state:} In this case, the $N$-particle 
wave function is given by \cite{Srednicki:1993im} (see also 
\cite{Plenio:2004he,Yarom:2004vp})
\beq
\Psi_0(x_1,\dots,x_n,x_{n+1},\dots,x_N) =
\le[\f{|\O|}{\p^N}\ri]^{1/4} 
\exp\le[ -\f{x^T\cdot \O \cdot x}{2} \ri] 
\eeq

\item {\it 1-particle state:} In this case, the $N$-particle 
wave function is given by \cite{Das:2005ah}
\ba
\Psi_{_{1}}(x_1 \dots x_N) &=&  \le| \f{\O}{4 \p^N}\ri|^{\f{1}{4}}
\sum_{i=1}^N a_i H_1 \le( k_{D i}^{\f{1}{4}} {x}_i\ri) 
\exp\le[ -\f{1}{2} k_{D i}^{\f{1}{2}}~{x}_i^2 \ri] \nn \\
& & \sqrt{2}~ \le( a^T K_D^{\f{1}{2}} {x} \ri)
 ~\psi_0\le( x_1,\dots,x_N \ri)
\ea 
where $a\mbox{'s are expansion coefficients}$ 
which satisfy the condition $a^T a=1$. For detailed discussion 
see \cite{Das:2005ah,Das:2006vx}.
\end{enumerate} 
%%%%%%%%%%%%%%%%%% Figure 
\begin{figure}[!htb]
\begin{center}
\includegraphics[height=120mm,width=135mm]{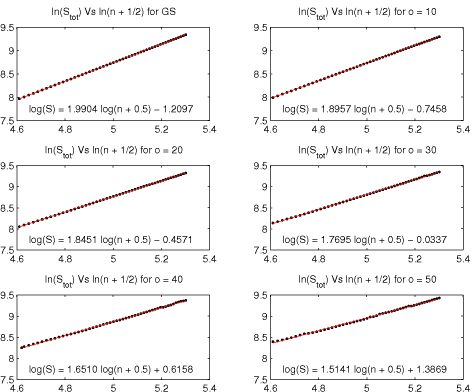}
\end{center}
\caption{The entanglement entropy for ground and 1-Particle states 
for $N = 300, n = 100 - 200, o = 10 - 50$}
\label{fig:3}
\end{figure}
%%%%%%%%%%%%%%%%%%%%%%%%%%%%%%

%%%%%%%%%%%%%%%%% subsection
\subsection{Density matrix}

For a general Eigen state, the physical operation of trace over region
2 ($n < N $ oscillators) [cf. Fig. (\ref{fig:2})] is given by
\br
\rho_{n} \le(x_{n + 1} \cdots x_{N}; x'_{n + 1} \cdots
x_{N}\ri) &= & \int \prod_{i=1}^n~dx_i~\Psi (x_1 \dots x_N) 
\Psi^\star (x_1 \dots x_n;x'_{n + 1} \cdots x'_N)  \nn
\er

For the two special choices which we discussed in the previous 
subsection, the above expression reduces to
\begin{enumerate}
\item {\it Vacuum state}: 
\beq
\rho_{_{GS}}= \prod_{i = 1}^n~ \rho_{2}(z_i, z_i') 
\eeq
which corresponds to product of $n$ 2-coupled harmonic oscillators with 
1 harmonic oscillator traced as discussed in Sec. (\ref{sec:2}).
\item {\it 1-Particle state}:
\beq
\rho_{_{ES}} = \rho_{_{GS}}  Tr (\L_A A^{-1})  \le[1 -\f{1}{2} \le( 
t^T\L_\c t + t'^T \L_\c t' \ri) + t^T\L_\b t' \ri]\, , \nn
\eeq
where 
\br
& & \L =  U^T~K_D^{\f{1}{4}}~a~a^T~K_D^{\f{1}{4}}~U \equiv
\le( 
\begin{array}{cc} 
\L_A & \L_B \\
\L_B^T & \L_C \end{array}
\ri)~; 
\Omega = U^{T} K_D^{1/2} U \equiv
\le( 
\begin{array}{cc} 
A & B \\
B^T & C 
\end{array}
\ri)~; \nn \\
%%%%%%%%%%%%%%%
& & t \equiv (t_1, \cdots t_{N -n}) =  (\vp_{n} \cdots \vp_{N}) \nn \\
%%%%%%
& & \L_\gamma = \f{2\L_B^T \le(A^{-1} B \ri) - 
B^T \le(A^{-1}\ri)^T \L_A A^{-1} B}
{Tr(\L_A A^{-1})} \nn \\
%%%%%%%
& & \L_\beta = \f{2\L_{_C} + B^T \le[A^{-1}\ri]^T \! \L_{_A} A^{-1} B 
- \L_{_B}^T A^{-1} B - B^T\!\le[A^{-1}\ri]^T\! \L_{_B}}
{Tr(\L_{_A} A^{-1})} 
\er
\end{enumerate}

As mentioned earlier, it is not possible to evaluate the
entanglement entropy analytically and we need to resort to numerical
computations. We use Matlab to evaluate the entropy and the numerical
error in our computation is less than $0.1 \%$.

In Fig. (\ref{fig:3}), we have plotted the entanglement entropy 
for the ground state and 1-particle state for $N = 300, 
n = 100 - 200, o = 10 - 50$. We see that for the 1-particle state, the entropy
scales as ${\mathcal A}^{\mu}$, with lesser $\mu$ for higher $o$. It is less
than unity for 
any $o > 0$ i. e., more the excitation, larger is deviation from area law
The area law does not seem to hold! This implies that the entanglement 
entropy depends on the choice of the quantum state of the field.

%%%%%%%%%%%%%%%%%%%%%%%%%%%%%% section
\section{Power law corrections to area}
\label{sec:4}

Given the above results, one may draw two distinct conclusions: first
that entanglement entropy is not robust and reject it as a possible
source of black-hole entropy.  Second, since entanglement entropy for
excited state scales as a lower power of area it is plausible that
when a generic state (consisting of a superposition of ground state and
excited state) is considered, corrections to the Bekenstein-Hawking
entropy will emerge. In order to determine which one is correct, it is
imperative to investigate various generalizations of the scenarios
considered in Ref. \cite{Das:2005ah}. To this end, in this section we
calculate the entanglement entropy of the mixed superposition of
vacuum and 1-particle state
\beq
\Psi = c_0 \Psi_0 + c_1 \Psi_1 \qquad {\rm where} 
\qquad {|c_0|^2 + |c_1|^2 = 1} \, .
\eeq
%
%%%%%%%%%%%%%%%%%%% Figure  
\begin{center}
\begin{figure}[!htb]
\includegraphics[angle=0,height=70mm,width=150mm]{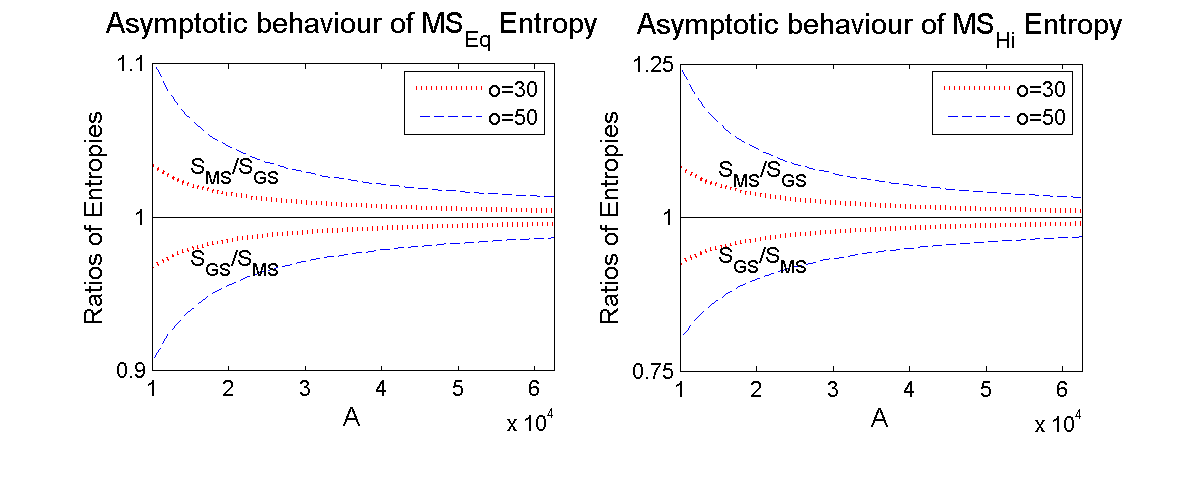}
\label{fig:4}
\caption{Plots of the relative entropy with the area for 
$N = 300, n = 100 - 200$ and $o - 30, 50$ where  
(i) $MS_{EQ}$ corresponds to $|c_0| = |c_1| = 0.7$
and (ii) $MS_{Hi}$ corresponds to $|c_0| = 0.5$, $|c_1| = 0.87$.}
\end{figure}
\end{center}
Following the procedure discussed in the previous section, it is
possible to obtain the entanglement entropy of the superposed
state. In Fig. (\ref{fig:4}) we have plotted the relative entropy [i.e
ratio of the superposed and ground state entanglement entropy] for
different values of $c_0$ and $c_1$. From Fig. (\ref{fig:4}) it is
easy to see that for large values of the horizon radius the
entanglement entropy of the superposed state approaches the ground
state entanglement entropy. In other words, for large horizon area,
the entanglement entropy of the superposed state scales as area of the
horizon.
%%%%%%%%%%%%%%%%%%% Figure 
\begin{center}
\begin{figure}[!hbt]
\includegraphics[angle=0,height=70mm,width=160mm]{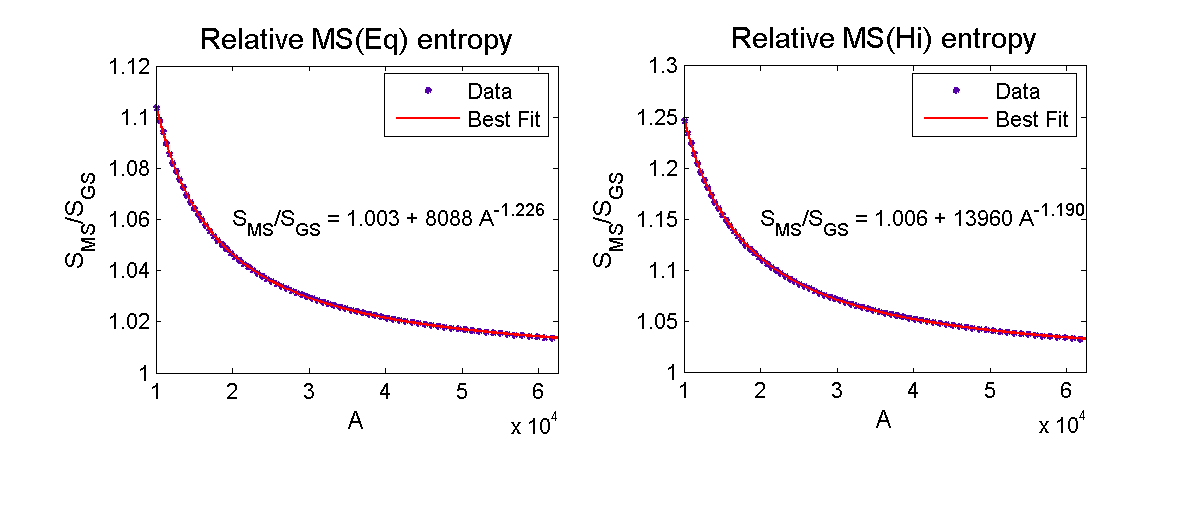}
\caption{The best fit the relative entropy vs ${\mathcal A}$ for 
(i) $MS_{EQ}$ corresponds to $|c_0| = |c_1| = 0.7$ and (ii)
$MS_{Hi}$ corresponds to $|c_0| = 0.5$, $|c_1| = 0.87$}
\label{fig:5}
\end{figure}
\end{center}
In order to know the behavior of the entanglement entropy for the
small horizon limit, in Fig. (\ref{fig:5}), we have obtained the best
numerical fit for the relative entropy. From these, we see that the
entanglement entropy of the superposed state is given by:
\beq
S = c_1 \, {\mathcal A} \,  \l(1  + \frac{c_2}{{\mathcal A}^{\beta}} \r)
\eeq
where $\beta > 0$ and $c_1, c_2$ are constants. This is the main
result of this paper and we would like to stress the following points:
(i) The second term in the above expression may be regarded as a power
law correction to the area law, resulting from entanglement, when the
wave-function of the field is chosen to be a superposition of ground
and excited states. (ii) It is important to note that the
correction term falls off rapidly with ${\mathcal A}$ (due to the
negative exponent) and in the large area limit (${\mathcal A}\gg 1$)
the area-law is recovered.  This lends further credence to
entanglement as a possible source of black hole entropy. (iii) The
correction term is more significant for greater excited state-ground
state mixing proportion $c_1$. For detailed discussion see
Refs. \cite{Das:2007sj,Das:2007mj}.

%%%%%%%%%%%%%%%%%%%%%Section
\section{Conclusions}
\label{sec:conc}

In this work, we have obtained power-law corrections to the
Bekenstein-Hawking entropy, treating the entanglement between scalar
field degrees of freedom inside and outside the horizon as a viable
source of black-hole entropy. We have shown that for small black hole
areas the area law is violated when the oscillator modes are in a
linear superposition of ground and excited states. We found that the
corrections to the area-law become increasingly significant as the
proportion of excited states in the superposed state increases.
Conversely, for large horizon areas, these corrections are relatively
small and the area-law is recovered.

Power law corrections to the Bekenstein- Hawking entropy has been
encountered earlier in other approaches to black hole entropy. For
instance, the Noether charge approach predicts a generic power law
correction to the Bekenstein-Hawking entropy \cite{Wald:1993nt}. 
For instance, using Noether charge as entropy, it was shown for 
5-dimensional Gauss-Bonnet gravity
\beq
{\mathcal S}_{GB}^{5D} = \frac{1}{16 \pi G} \int d^5 x \sqrt{-g} \l[R +
\alpha_{GB} \l(R^2 - 4 R_{ab} R^{ab} + R_{abcd} R^{abcd} \r) \r]
\quad \alpha_{GB} \mbox{~is a constant} \, ,
\eeq
that the entropy corresponding to a Killing horizon is \cite{Myers:1988ze}
\beq
S_{_{\rm NC}}^{5D} = \frac{A}{4}  \l(1 + \frac{c_3}{A^{2/3}} \r)  \, .
\eeq
Although both these approaches lead to power-law corrections, there
are couple of crucial differences between the two: (i) In the case of
Noether charge, the power-law corrections to the Bekenstein-Hawking
entropy arises due to the higher-derivative terms in the gravity
action. However, the corrections we have derived is also valid for
Einstein-Hilbert action. (ii) The Noether charge entropy does not
identify which degrees of freedom contribute most to the entropy. In
our case, we have shown that the maximum entropy contribution for the
area-law come close to the horizon while significant contribution to
the power-law corrections arise from the degrees of freedom far from
the horizon \cite{Das:2007sj,Das:2007mj}.

%\section*{Acknowledgments}
\ack
SS is supported by the Marie Curie Incoming International Grant
IIF-2006-039205.

\appendix

\section{Scalar fields in Schwarzschild space-time}
In this section, we generalize the framework of our calculations,
so that they are applicable to black hole space-times. We start
with the Schwarzschild metric with a horizon at $r=r_0$: 
\be
ds^2 = -f(r) dt^2 + \f{dr}{f(r)} + r^2 d\O^2 ~,~~f(r)= 1 - \f{r_0}{r} 
\la{sch1}
\ee
and define the following coordinate transformation from $(t,r)$ 
to $(\tau,R)$ coordinates:
\ba
\tau = t + r_0 \le[ \ln\le( \f{1-\sqrt{r/r_0} }{1+\sqrt{r/r_0}} \ri) 
+ 2\sqrt{\f{r}{r_0}}  \ri]~,~
R = \tau + \f{2~r^{\f{3}{2}}}{3\sqrt{r_0}} ~,~
r = \le[ \f{3}{2} \le( R-\t\ri) \ri]^{\f{2}{3} } r_0^{\f{1}{3} }~,
\la{transfn1}
\ea
such that (\ref{sch1}) transforms to the following metric, in 
{\it Lemaitre coordinates}:  
\be
ds^2= -d\t^2 + 
\f{dR^2}{\le[ \f{3}{2r_0} \le( R-\t\ri) \ri]^{\f{2}{3}} }
+ \le[ \f{3}{2r_0} \le( R-\t\ri) \ri]^{\f{4}{3}}r_0^{\f{2}{3} }~d\O^2 ~. 
\ee
The Hamiltonian for a free scalar field in the above space-time can be
written as 
\footnote{the current analysis is done for $l=0$. For generalization 
see Ref. \cite{Das:2007mj}.}:
\be
H(\t)= \f{1}{2} \int_\tau^\infty dR  
\le[
\f{2\p(\t,R)^2}{3(R-\t)} + \f{3}{2}~r~(R-\t) \le( \pa_R \phi(\t,R) \ri)^2  
\ri]~.
\la{schham}
\ee
Next, choose a {\it fixed} Lemaitre time, say $\t=0$ and perform the
following field redefinitions:
\be
\p(r) = \sqrt{r} \p_1(r) ~,~~\phi (r) = \f{\phi_1(r)}{r}~.  
\ee
Then, at that {\it fixed time}, the 
Hamiltonian (\ref{schham}) transforms to: 
\be
H(0) = \f{1}{2} \int_0^\infty dr
\le[
\p_1(r)^2 + r^2 \le( \pa_r \f{\phi_1}{r} \ri)^2
\ri]~.
\ee
Note that this is identical to the Hamiltonian in flat space-time,
Eq.(\ref{ham2}). Hence, all the analysis in
Secs. (\ref{sec:2},\ref{sec:3}) will go through for a 
black-hole space-time in a fixed Lemaitre time.

\end{document}